\documentclass{article}
\usepackage{graphicx}
\begin{document}
\bibliographystyle{num}
\title{Superfluidity with dressed nucleons}
\baselineskip=1. \baselineskip

\author{     P. Bo\.{z}ek\footnote{Electronic
address~:
piotr.bozek@ifj.edu.pl}\\
Institute of Nuclear Physics, Pl-31-342 Cracow, Poland}

\date{\today}

\maketitle

\vskip .3cm


\vskip .3cm

\begin{abstract}

The gap equation with dressed propagators is solved
 in symmetric nuclear matter. Nucleon self-energies are obtained within the
self-consistent in medium $T$ matrix  approximation. The off-shell gap
 equation is compared to an effective quasiparticle gap equation with reduced 
interaction. At normal density, we find
 a reduction of the superfluid gap from
 $6.5{\rm  MeV}$ to $0.45{\rm  MeV}$ when
self-energy effects are included.
\end{abstract}

{superfluidity, nuclear matter

\vskip .4cm

{\bf  21.65.+f, 24.10.Cn, 26.60.+c}

\vskip .4cm

The importance of
pairing correlation in nuclear systems was realized very early
\cite{bohrpr110}. In finite nuclei, pairing effects 
are known in the mass systematics of nuclei 
and the properties of deformed nuclei.
In extended systems, nuclear pairing is expected to occur in the dense
matter inside the neutron stars \cite{migdal1,baymNature}. 
Quantitative description of pairing correlations starting from bare
nucleon-nucleon interactions is still limited.
In medium effects, going beyond simple gap equation with free interaction, 
are important \cite{clarkpol,Wambach:1993ik,schulzepol,pol2,Bozek:2000fn,Baldo:2000zf,Lombardo:2001vp,karz,milano}. The use of the  induced interaction
in the gap equation was studied for neutron matter leading to a significant
reduction of the pairing strength 
\cite{clarkpol,Wambach:1993ik,schulzepol,pol2}.
Another effect is due to the dressing of nucleons in an 
interacting system, leading to a
modification of the density of states and of the effective energy gap
\cite{Bozek:2000fn,Baldo:2000zf,Lombardo:2001vp}. From the
Bardeen-Cooper-Schrieffer  result \cite{bcs}, it is obvious that 
self-energy effects modifying the effective mass change the
value of the superfluid gap. Furthermore, 
the reduced quasiparticle strength at the Fermi
surface leads  to a modification of the effective pairing interaction
\cite{Bozek:2000fn,Baldo:2000zf,Lombardo:2001vp}.

Recently self-consistent in medium $T$ matrix was calculated 
for nuclear matter \cite{di1,Bozek:1998su,Dickhoff:1999yi,Dewulf:2000jg,Bozek:2001tz,Alm:1996ps}.
In this approach ladder diagrams with dressed 
particle-particle and hole-hole propagators are summed 
\begin{eqnarray}
\label{teq}
& & \langle{\bf p}|T({\bf P},\Omega)|{\bf p}^{'}\rangle = V({\bf p},{\bf p}^{'})
\nonumber \\ & & + 
 \int\frac{d\omega_1}{2\pi}\int\frac{d\omega_2}{2\pi}
\int\frac{d^3q}{(2 \pi)^3} V({\bf p},{\bf q})
\frac{\left[1-f(\omega_1)-f(\omega_2)\right]}
{\Omega-\omega_1-\omega_2+i\epsilon} \nonumber \\ & &
A(p_1,\omega_{1})A(p_2,\omega_{2})
 \langle{\bf q}|T({\bf P},\Omega)
|{\bf p}^{'}\rangle
\end{eqnarray}
where ${\bf p_{1,2}}={\bf P}/2\pm {\bf q}$, $f(\omega)$ is the Fermi 
distribution, and 
\begin{equation}
\label{spectralf}
A(p,\omega)=\frac{-2 {\rm Im}\Sigma(p,\omega)}{\left[\omega-p^2/2m 
-{\rm Re}\Sigma(p,\omega)\right]^2 +{\rm Im}\Sigma(p,\omega)^2}
\end{equation}
is the self-consistent spectral function of the nucleon.
The  imaginary part of the corresponding retarded self-energy can be obtained
from
\begin{eqnarray}
\label{imags}
& & 
{\rm Im}
\Sigma(p,\omega) =\int\frac{d\omega_1}{2 \pi}\int \frac{d^3k}{(2 \pi)^3}
A(k,\omega_1) \nonumber \\ 
& & \langle({\bf p}-{\bf k})/2|{\rm Im}T({\bf p}
+{\bf k},\omega+\omega_1)|({\bf p}-{\bf k})/2\rangle_A
 \left[ f(\omega_1)+b(\omega+\omega_1) \right] \ ,
\end{eqnarray}
 $b(\omega)$
is the Bose distribution.
The real part of the self-energy is related to ${\rm Im} \Sigma$
by a dispersion relation
\begin{equation}
\label{reals}
{\rm Re}\Sigma(p,\omega)= \Sigma_{HF}(p) + {\rm P} \int \frac{d\omega^{'}}
{\pi} \frac{{\rm Im}\Sigma(p,\omega^{'})}{\omega^{'}-\omega}
\end{equation}
with $\Sigma_{HF}(p)$ the Hartree-Fock self-energy.
Eqs. (\ref{teq}), (\ref{imags}),   (\ref{spectralf}), and (\ref{reals})
are solved  iteratively. 
Numerical calculations with off-shell propagators are very complex;
recently, the self-consistent $T$ matrix approximation scheme was solved for a
realistic interaction with several partial waves \cite{Bozek:2002em}.
It is the aim of the present work to analyze the effects of the self-consistent
dressing of fermion propagators on nucleon superfluidity for such a 
realistic interaction. For the case of the symmetric nuclear matter, the
dominant pairing correlations appear in the $^3S_1-^3D_1$ partial wave with
the value of the superfluid gap of the order of 
several MeV's \cite{Vonderfecht:1991,baldosd,heavyiond}.
Such a large value of the superfluid gap is not seen in nuclear mass 
systematics for $N\simeq Z$ nuclei. Although the
existence
of 
nuclear pairing in the $^1S_0$ channel is well established in the data,
the calculation of the corresponding
 pairing force from the bare interaction is not available.
The strength of the  $^1S_0$  pairing force in a simple mean-field
approach is small, unlike for the deuteron channel. Pairing gaps
observed in finite nuclei result from bulk and surface pairing
interactions 
\cite{karz,milano}. The first mechanism should be dominant for the
$^3S_1-^3D_1$ channel if the large value of the BCS gap is taken,
which would result in an
 unrealistically large value of the proton-neutron gap. No such
paradox exist for the $^1S_0$ channel, where the BCS gap in nuclear
matter is small, which means that the pairing force in finite nuclei
has (at least partly) a different origin.
It is also important to clarify the role of dominant pairing
correlations for nuclear matter calculations, especially in view of
recent suggestions that normal nuclear matter is not superfluid 
\cite{Dewulf:2000jg,Frick:2001jp,Dickhoff:1999yi} 
due to self-energy corrections.
 In the following, we demonstrate
explicitly that self-energy corrections reduce the value of the superfluid
gap to a small, but nonzero, value 
in symmetric nuclear matter at saturation density $\rho_0$.

From the Thouless criterion on the 
self-consistent $T$ matrix  at finite temperature, 
 it is known that self-consistent dressing of nucleons reduces 
significantly the critical temperature \cite{Bozek:1998su}.
Calculation performed in the normal phase of cold nuclear matter suggest a
strong suppression of pairing correlations due nucleon dressing
\cite{Dickhoff:1999yi,Bozek:2001tz}.  However, no explicit calculation of
superfluid properties with  dressed propagators and realistic 
interaction is available.
In Ref. \cite{Bozek:1999rv} the gap equation with 
off-shell propagators
\begin{equation}
\label{gapfull}
\Delta(p)=\int\frac{d \omega d\omega^{'} d^3k}{(2\pi)^5} A(k,\omega) 
A_s(k,\omega)\frac{\left[1-f(\omega)-f(\omega^{'})\right]}{-\omega-\omega^{'}}
V(p,k)\Delta(k)
\end{equation}
 was solved for a simple interaction.
In the above equation $A_s(p,\omega)$ denotes the spectral function of the
nucleon, including the diagonal self-energy $\Sigma(p,\omega)$ 
(obtained in the $T$ matrix
 approximation) and the off-diagonal self-energy $\Delta(p)$ obtained from
Eq. (\ref{gapfull}) itself; 
$A(p,\omega)$ is the spectral function of the nucleon dressed
with the  diagonal self-energy only (\ref{spectralf}). 
It was found that the superfluid gap for dressed nucleons
 is reduced by a factor $2-3$ in comparison to the result of the quasiparticle 
gap equation. 

Self-energy effects can be effectively taken into account in a quasiparticle
gap equation \cite{Baldo:2000zf,Bozek:2000fn,Lombardo:2001vp}.
The first consequence of dispersive
self-energy corrections to the propagator is that the superfluid energy gap 
is  not the off-diagonal
self-energy $\Delta(p)$ but \cite{Bozek:2000fn}
\begin{equation} 
\hat{\Delta}(p)=Z_p\Delta(p)
\end{equation}
 where 
\begin{equation}
Z_p=\left(1-\frac{\partial \Sigma(p,\omega_p)}{\partial
    \omega}|_{\omega=\omega_p}\right)^{-1} \ .
\end{equation}
It means that if the position of the quasiparticle peak in $A(p,\omega)$
 is given by~:
\begin{equation}
\omega_p=\frac{p^2}{ m} +{\rm Re}{\Sigma}(p,\omega_p) \ ,
\end{equation}
the poles of the propagator in the superfluid are located approximately at
\begin{eqnarray}
\omega=\pm E_p&=&\pm\sqrt{(\omega_p-\mu)^2+\hat{\Delta}(p)^2} \nonumber \\
& =&\pm\sqrt{(\omega_p-\mu)^2+Z_p^2{\Delta(p)}^2} \ .
\end{eqnarray}
The quasiparticle strength at the poles $\pm E_p$ of the superfluid propagator
can be expressed, to a very good accuracy, 
by the quasiparticle renormalization strength $Z_p$
of the pole of the normal propagator \cite{Bozek:2000fn}.
The quasiparticle effective gap equation with dressed propagators is
\cite{Bozek:2000fn,Lombardo:2001vp}
\begin{equation}
\label{gapz}
\Delta(p)=-\int\frac{d^3k}{(2\pi)^3}Z_k^2 V(p,k)\frac{\left[1-2f(E_k)\right]}
{2E_k}\Delta(k) \ .
\end{equation}
The effective interaction in the gap equation is renormalized 
by the factor $Z_k^2$ describing the reduction of the density of states at the
Fermi surface (Eq. (\ref{gapz}) can also be written as an equation for
the energy gap $\hat{\Delta}(p)$ with reduced interaction $Z_pZ_kV(p,k)$). 
The superfluid gaps obtained from
the gap equation (\ref{gapfull}) or the corresponding effective quasiparticle
gap equation (\ref{gapz}) are much smaller than the one
obtained from  the usual gap equation with the  bare interaction~:
\begin{equation}
\label{gapbcs}
\Delta(p)=-\int\frac{d^3k}{(2\pi)^3} V(p,k)\frac{\left[1-2f(E_k)\right]}
{2\sqrt{(\omega_k-\mu)^2+\Delta(k)^2}}\Delta(k) \ .
\end{equation}

In the present work,
we  calculate the diagonal self-energy in the $T$ matrix approximation
for symmetric nuclear matter in the range of densities $.1-2\rho_0$.
As a first approximation, a normal state of  nuclear matter is assumed.
Certainly, if a superfluid phase with a significant gap is present
 the approach
must be modified \cite{Bozek:2001nx}. However, to get a first estimate for the
value of the gap with dressed propagators we can take diagonal-self energies
(single-particle potential and scattering width) corresponding to the normal
phase. It is a reasonable approximation around the saturation density where
the resulting superfluid gap is small.
It is analogous to  approaches using Brueckner-Hartree-Fock single particle spectrum or 
variational wave functions calculated in the normal phase as input 
for the gap equation. We use the separable Paris interaction for the 
calculation of the spectral functions \cite{Bozek:2002em}.
From the real part of the self-energy,
 we obtain the single-particle energies $\omega_p$ and
the quasiparticle strength $Z_p$.

In Fig \ref{gapfig} are presented the values of the superfluid energy gap
$Z_{p_F}\Delta(p_F)$ at the Fermi momentum 
obtained in the different approximations. We calculate the pairing gap
in the $^3S_1-^3D_1$ channel for the Paris interaction using an angle averaged
gap for the off-diagonal self-energy.
The largest value of the superfluid gap comes from the gap equation with the
bare interaction (\ref{gapbcs}) where the sole influence of the nuclear
medium comes trough the modification of  single-particle energies.
On the other hand, the gap equation with dressed propagators (\ref{gapfull}) 
leads to much
smaller values of the superfluid gap.
We find the maximal value of the superfluid gap $\hat{\Delta}(p_F)\simeq 2.2
{\rm MeV}$ at $.5\rho_0$. At the saturation density $\hat{\Delta}(p_F)= 0.45
{\rm MeV}$. The last value is significantly smaller than the corresponding
BCS gap (\ref{gapbcs}) $\Delta(p_F)=6.5{\rm MeV}$. The reasons are twofold.
First, the effective energy gap is $Z_p\Delta(p)$ instead of $\Delta(p)$
due to the dispersive self-energy \cite{Bozek:2000fn}. Second, the spectral
functions for fully dressed nucleons ($A(p,\omega)$ and $A_s(p,\omega)$) 
give an effective reduction of the density of states at the Fermi surface
\cite{Bozek:2000fn}. Both effects are taken into account in
the effective gap equation with reduced interactions (\ref{gapz}).
Indeed, the superfluid gap obtained from the effective gap 
equation (\ref{gapz}) 
is close to the solution of the full equation (\ref{gapfull}).

In Fig. \ref{kernelfig} are compared the kernels of the full gap equation 
(\ref{gapfull}) 
\begin{equation}
\label{kernelfull}
\frac{1}{2\langle E_k\rangle} = \int 
\frac{d \omega d\omega^{'}}{(2\pi)^2} A(k,\omega) 
A_s(k,\omega)\frac{\left[1-f(\omega)-f(\omega^{'})\right]}
{\omega+\omega^{'}}
\end{equation}
and
of the effective gap equation (\ref{gapz})
\begin{equation}
\label{kernelz}
\frac{1}{2 E_k} = \frac{Z_k^2}{\sqrt{(\omega_k-\mu)^2+\hat{\Delta}(k)^2}} \ .
\end{equation}
As expected, close to the Fermi momentum the kernel appearing in the gap
equation with full spectral functions can be approximated by the 
renormalization of the quasiparticle poles in the quasiparticle gap equation.
The momentum integration in the $^3S_1-^3D_1$ gap equation cannot be
restricted to the vicinity of the Fermi momentum  only.
Differences between the kernels of the gap equations at low
momenta lead to small differences in the resulting gaps (Fig. \ref{gapfig}).
It must be stressed, however, that to a reasonable accuracy, the 
effective kernel (\ref{kernelz}) is a good approximation of the kernel of the
full gap equation (\ref{kernelfull}) and describes the mechanism of the
reduction of the superfluid gap by nucleon dressing. 
It means that the background part of the spectral function does not modify
significantly the effective interaction between quasiparticles 
 in the gap equation. 

Similar effects are visible in the temperature dependence of the superfluid
gap. The largest gap and the largest critical temperature is obtained from
the gap equation with the bare interaction (Fig. \ref{tempfig}).
The gap equation with dressed nucleons (\ref{gapfull}) and the effective
 gap equation (\ref{gapz}) give much smaller values 
for the critical temperature.
The single-particle potential is weakly dependent on the temperature 
\cite{Bozek:2002em}, and the gap closure is due in all the approaches to the 
phase space factor $\left[1-f(\omega)-f(\omega^{'})\right]$. 
At the critical temperature the effective gap equation (\ref{gapz})
corresponds to the Thouless condition for the quasiparticle 
$T$ matrix with a reduced
interaction $Z_k^2 V(p,k)$ \cite{Bozek:2000fn}.

We calculate the superfluid gap for the deuteron channel 
 with off-shell propagators. The spectral functions serving as input for the
 gap equation are obtained in the self-consistent $T$ matrix approximation.
The most important result is the strong reduction of the superfluid energy gap
in symmetric nuclear matter in comparison to results 
obtained with bare interactions. The superfluid gap is $0.45{\rm MeV}$ 
at the saturation density, when self-energy effects are included.
The actual value of the superfluid gap in symmetric nuclear matter is 
influenced also by vertex corrections (the induced interaction)
\cite{clarkpol,Wambach:1993ik,schulzepol,pol2,Schwenk:2001hg,rgfermi,fermirmp}.

The strong reduction of the gap that we find is important in
 two respects. First, such a small value of the superfluid energy gap
 in nuclear matter justifies
 standard approaches to the nuclear matter
 problem
which neglect the superfluid transition.
Second, the small value of the neutron-proton pairing gap in nuclear
 matter is 
 compatible with common
 inferred  from the properties of finite nuclei. 
The value, or even the presence, of the a neutron-proton gap in finite
 nuclei is still a matter of debate 
\cite{Satula:1997dc,Ropke:1999an,Kelsall:2002du,Jenkins:2002qe};
but a large value of the neutron-proton gap, as obtained form a simple
 BCS
gap equation with bare interaction, is excluded by the data.
Since we find a small value of the $^3S_1-^3D_1$ gap, the paradox is
 resolved. However, it is even more difficult to make definite
 predictions on the value of the energy gap in finite nuclei than in
 nuclear matter. Besides  vertex
 corrections other mechanism of the pairing interactions in finite
 nuclei are possible, that cannot be treated in a local density approximation.

As a byproduct, we study the 
quasiparticle gap equation with effective interactions. We find a reasonable
 agreement with the results obtained with complete spectral functions.
Therefore,
the effective interaction $Z_k^2 V(p,k)$ is a good starting point for 
the study of other many-body effects,  such as the role of the 
induced interaction 
\cite{clarkpol,Wambach:1993ik,schulzepol,pol2,Schwenk:2001hg,rgfermi,fermirmp},
 for the
 pairing channel. In neutron matter a similar reduction of the superfluid gap
 is expected. The effect would be smaller since in medium dispersive
 effects are weaker \cite{Baldo:2000zf,Lombardo:2001vp}.

\vskip .3cm
This work was partly supported by the KBN
under Grant No. 2P03B02019.

\bibliography{../mojbib}

\begin{thebibliography}{10}
\expandafter\ifx\csname url\endcsname\relax
  \def\url#1{\texttt{#1}}\fi
\expandafter\ifx\csname urlprefix\endcsname\relax\def\urlprefix{URL }\fi

\bibitem{bohrpr110}
A.~Bohr, B.~R. Mottelson, D.~Pines, Phys. Rev., {\bf 100} (1958) 936.

\bibitem{migdal1}
A.~B. Migdal, Zh. Eksp. Teor. Fiz, {\bf 37} (1959) 249.

\bibitem{baymNature}
G.~Baym, C.~Pethick, D.~Pines, Nature, {\bf 224} (1969) 673.

\bibitem{clarkpol}
J.~Clark, C.-G. {K\"allman}, C.-H. Yang, D.~Chakkalakal, Phys. Lett., {\bf B61}
  (1976) 331.

\bibitem{Wambach:1993ik}
J.~Wambach, T.~L. Ainsworth, D.~Pines, Nucl. Phys., {\bf A555} (1993) 128.

\bibitem{schulzepol}
H.-J. Schulze, J.~Cugnon, A.~Lejeune, M.~Baldo, U.~Lombardo, Phys. Lett., {\bf
  B375} (1996) 1.

\bibitem{pol2}
H.-J. Schulze, A.~Polls, A.~Ramos, Phys. Rev., {\bf C63} (2001) 044310.

\bibitem{Bozek:2000fn}
P.~Bo\.zek, Phys. Rev., {\bf C62} (2000) 054316.

\bibitem{Baldo:2000zf}
M.~Baldo, A.~Grasso, Phys. Lett., {\bf B485} (2000) 115.

\bibitem{Lombardo:2001vp}
U.~Lombardo, P.~Schuck, W.~Zuo, Phys. Rev., {\bf C64} (2001) 021301.

\bibitem{karz}
A.~V. Avdeenkov, S.~P. Kamerdzhiev, Yad. Fiz., {\bf 62N} (1999) 610.

\bibitem{milano}
J.~Terasaki, F.~Barranco, P.~F. Bortignon, R.~A. Broglia, E.~Vigezzi, Nucl.
  Phys., {\bf A697} (2002) 127.

\bibitem{bcs}
J.~Bardeen, L.~N. Cooper, J.~R. Schrieffer, Phys. Rev., {\bf 108} (1957) 1175.

\bibitem{di1}
W.~H. Dickhoff, Phys. Rev., {\bf C58} (1998) 2807.

\bibitem{Bozek:1998su}
P.~Bo\.zek, Phys. Rev., {\bf C59} (1999) 2619.

\bibitem{Dickhoff:1999yi}
W.~H. Dickhoff, C.~C. Gearhart, E.~P. Roth, A.~Polls, A.~Ramos, Phys. Rev.,
  {\bf C60} (1999) 064319.

\bibitem{Dewulf:2000jg}
Y.~Dewulf, D.~Van~Neck, M.~Waroquier, Phys. Lett., {\bf B510} (2001) 89.

\bibitem{Bozek:2001tz}
P.~Bo\.zek, P.~Czerski, Eur. Phys. J., {\bf A11} (2001) 271.

\bibitem{Alm:1996ps}
T.~Alm, G.~{R\"opke}, A.~Schnell, N.~H. Kwong, H.~S. Kohler, Phys. Rev., {\bf
  C53} (1996) 2181.

\bibitem{Bozek:2002em}
P.~Bo\.zek, Phys. Rev., {\bf C65} (2002) 054306.

\bibitem{Vonderfecht:1991}
B.~Vonderfecht, C.~Gerhart, W.~Dickhoff, A.~Polls, A.~Ramos, Phys. Lett., {\bf
  B253} (1991) 1.

\bibitem{baldosd}
M.~Baldo, I.~Bombaci, U.~Lombardo, Phys. Lett., {\bf B283} (1992) 8.

\bibitem{heavyiond}
M.~Baldo, U.~Lombardo, P.~Schuck, Phys. Rev., {\bf C52} (1995) 975.

\bibitem{Frick:2001jp}
T.~Frick, K.~Gad, H.~Muther, P.~Czerski, Phys. Rev., {\bf C65} (2002) 034321.

\bibitem{Bozek:1999rv}
P.~Bo\.zek, Nucl. Phys., {\bf A657} (1999) 187.

\bibitem{Bozek:2001nx}
P.~Bozek, Phys. Rev., {\bf C65} (2002) 034327.

\bibitem{Schwenk:2001hg}
A.~Schwenk, G.~E. Brown, B.~Friman, Nucl. Phys., {\bf A703} (2002) 745.

\bibitem{rgfermi}
S.~K. Bogner, A.~Schwenk, T.~T.~S. Kuo, G.~E. Brown, nucl-th/0111042 (2001).

\bibitem{fermirmp}
R.~Shankar, Rev. Mod. Phys., {\bf 66} (1994) 129.

\bibitem{Satula:1997dc}
W.~Satula, R.~Wyss, Phys. Lett., {\bf B393} (1997) 1.

\bibitem{Ropke:1999an}
G.~{R\"opke}, A.~Schnell, P.~Schuck, U.~Lombardo, Phys. Rev., {\bf C61} (2000)
  024306.

\bibitem{Kelsall:2002du}
N.~S. Kelsall, {\it et~al.}, Phys. Rev., {\bf C65} (2002) 044331.

\bibitem{Jenkins:2002qe}
D.~G. Jenkins, {\it et~al.}, Phys. Rev., {\bf C65} (2002) 064307.

\end{thebibliography}

\newpage

\begin{figure}

\centering
\includegraphics[width=0.7\textwidth]{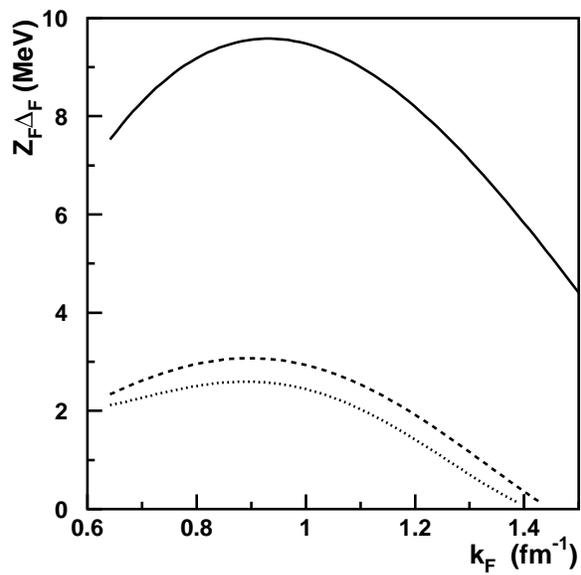}
\caption{The value of the superfluid energy gap at the Fermi momentum 
$\Delta_{p_F}Z_{p_F}$ as a function
  of the density
for the BCS gap equation [Eq. (\ref{gapbcs})] (solid line), for the gap equation with dressed
propagators  [Eq. (\ref{gapfull})] (dotted line), 
and for the effective gap equation [Eq. (\ref{gapz})] (dashed line).}
\label{gapfig}
\end{figure}

\begin{figure}

\centering
\includegraphics[width=0.7\textwidth]{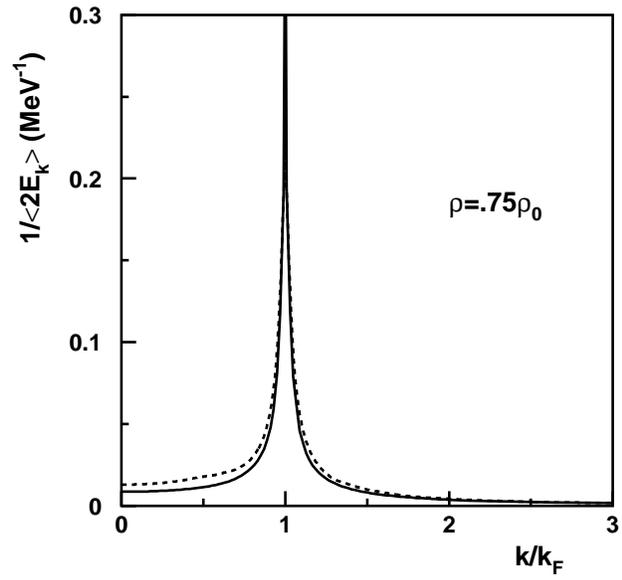}
\caption{The kernel of the full gap equation [Eq. (\ref{kernelfull})] (solid line)
and of the effective gap equation [Eq. (\ref{kernelz})] (dashed line), at $\rho=.75\rho_0$.}
\label{kernelfig}
\end{figure}

\begin{figure}

\centering
\includegraphics[width=0.7\textwidth]{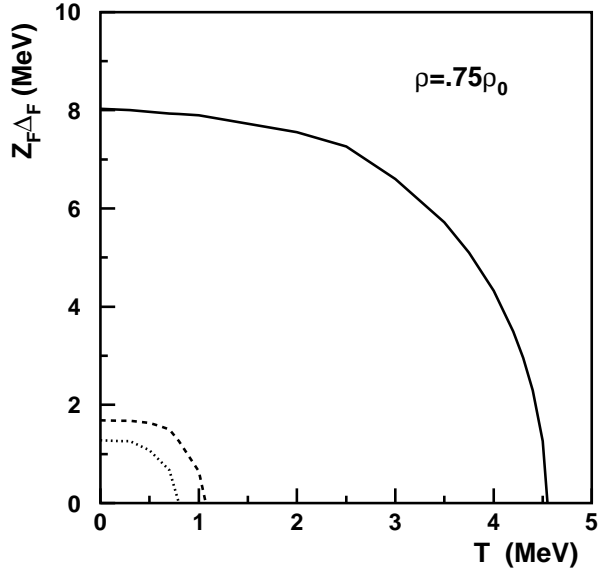}
\caption{The value of the superfluid energy gap at the Fermi momentum 
$\Delta_{p_F}Z_{p_F}$ as a function
  of the temperature 
for the BCS gap equation [Eq. (\ref{gapbcs})] 
(solid line), for the gap equation with dressed
propagators  [Eq. (\ref{gapfull})] (dotted line), 
and for the effective gap equation [Eq. (\ref{gapz})] (dashed line), at
  $\rho=.75\rho_0$.}
\label{tempfig}
\end{figure}

\end{document}